\documentclass[a4paper,11pt]{article}
\usepackage{jcappub} 
\usepackage{lineno}

\usepackage[utf8]{inputenc}
\usepackage[T1]{fontenc}
\usepackage{comment}
\usepackage{bm}
\usepackage{mathtools,amsmath,amssymb,amsfonts,mathrsfs,eucal,graphicx,tensor,csquotes,accents,commath,chngcntr,siunitx}

\newcommand{\dd}{\mathrm{d}}

\title{\boldmath Restoring Causality in Higher Curvature Gravity}

\author[a]{Jose D. Edelstein,}
\author[b,c]{Rajes Ghosh,}
\author[d]{Alok Laddha,}
\author[e]{Sudipta Sarkar}

\affiliation[a]{Departamento de F\'\i sica de Part\'\i culas $\&$ Instituto Galego de F\'\i sica de Altas Enerx\'\i as (IGFAE), Universidade de Santiago de Compostela, E-15782 Santiago de Compostela, Spain}

\affiliation[b]{William H. Miller III Department of Physics \& Astronomy, Johns Hopkins University,
3400 North Charles Street, Baltimore, MD 21218, USA}

\affiliation[c]{International Centre for Theoretical Sciences, Tata Institute of Fundamental Research, Bangalore 560089, India.}

\affiliation[d]{Chennai Mathematical institute, Chennai, Tamil Nadu 603103 India.}

\affiliation[e]{Indian Institute of Technology, Gandhinagar, Gujarat 382355, India.}

\emailAdd{jose.edelstein@usc.es}
\emailAdd{rajes.ghosh@icts.res.in}
\emailAdd{aladdha@cmi.ac.in}
\emailAdd{sudiptas@iitgn.ac.in}

\abstract{Incorporating higher curvature terms into gravity theories modifies the classical field equations, potentially leading to theoretical issues like Shapiro time advancements that violate the Camanho, Edelstein, Maldacena, and Zhiboedov (CEMZ) causality criterion. We explore this criterion within the context of Generalised Quadratic Gravity (GQG), a higher curvature theory with a distinct graviton three-point coupling from General Relativity (GR). By constructing an exact shock wave solution of GQG and calculating the Shapiro time shift for a massless probe graviton, we demonstrate that it can remain strictly positive within a classically allowed parameter space of couplings, ensuring the theory's adherence to the CEMZ criterion, due to the presence of perturbative ghosts. This finding indicates that GQG can offer a causal extension beyond GR, paving the way for further exploration into the consistency of classical higher curvature gravity theories.}

\begin{document}
\maketitle
\flushbottom

\section{Introduction}
General Relativity (GR) has been overwhelmingly successful in explaining gravitational dynamics in weak field regimes~\cite{Will:2014kxa}. However, there is still insufficient evidence to confirm its validity in strong gravitational fields, where potential deviations from GR will be most pronounced~\cite{Berti:2015itd, Berti:2019tcy}. Such deviations often manifest as higher curvature terms supplementing the Einstein-Hilbert action, introducing new dimensionful couplings relevant at certain classical length scales. Higher-curvature gravity theories also arise as an extension of GR, motivated by the quest for a more complete theory of gravity that remains well-behaved at high energies. These theories have been extensively studied in the context of black hole physics, gravitational waves and cosmology. 

However, the inclusion of such higher curvature terms typically produces modifications in classical field equations, introducing new features that may result in theoretical pathologies. For example, higher curvature theories often lead to higher-order field equations with Ostrogradsky-like instabilities, perturbative ghosts, and tachyonic modes~\cite{Stelle:1977ry, Audretsch:1993kp, Salvio:2018crh}. Nevertheless, there exist examples of theories that can evade some of these pathologies~\cite{Padmanabhan:2013xyr, Camanho:2013pda, Bueno:2016ypa, Bueno:2019ltp}, including those arising from the ghost modes~\cite{Deffayet:2023wdg, Held:2023aap, Figueras:2024bba, Held:2025ckb}. However, such constructions typically rely on additional structures, like constraints, gauge symmetries, or integrability conditions, which render the ghost modes non-propagating and physically benign. 

At the quantum level, the situation becomes significantly more severe. As emphasized in earlier works ~\cite{Cline:2003gs} (see also Refs.~\cite{Antoniadis:1986tu, Smilga:2005gb}), theories containing propagating ghost degrees of freedom generally suffer from catastrophic pathologies associated with an unbounded Hamiltonian and vacuum instability. To our knowledge, there is currently no universally accepted mechanism that resolves such quantum instabilities. In view of these considerations, the question of classical and quantum stability in gravitational theories with ghost degrees of freedom remains an important open problem. This makes it imperative to continue exploring such theories to better understand the implications of ghost modes and to develop criteria that can systematically identify theoretically ``healthy''  gravity theories.

A key idea driving the recent progress in classifying healthy gravitational theories is based on the so-called \textit{causality constraint} proposed by Camanho, Edelstein, Maldacena, and Zhiboedov (CEMZ) \cite{CEMZ}. They demonstrated that theories with negative Shapiro time shifts for a probe graviton as it crosses a shock wave are acausal. This criterion can also be restated in terms of the graviton $2-2$ scattering amplitudes in the eikonal limit. In particular, any theory having a massless graviton as the only propagating mode, such as Einstein-Gauss-Bonnet (EGB) gravity, is ruled out if they deform the graviton three-point coupling relative to its GR value~\cite{CEMZ}.

Recently, CEMZ's idea has been extended to a class of higher derivative theories including Quadratic Gravity (QG) where the perturbative Lagrangian around flat spacetime has the same three-point coupling as GR, but the propagating modes include a massive spin-2 ghost besides the massless graviton~\cite{EGLS}. The work showed that these theories are also consistent with the CEMZ criterion, albeit in a non-trivial way: the ghost massive graviton contributes to the Shapiro delay without flipping its sign. However, the results of Refs.~\cite{CEMZ,EGLS} do not apply to theories that are not equivalent to EGB theory or QG under field redefinitions of the metric. These distinct theories, which our present work addresses in terms of the causality issue, are characterized by different three-point graviton couplings relative to GR and additional propagating modes (like ghosts) other than the massless gravitons. For concreteness, we focus on the simplest prototype, namely \textit{Generalised Quadratic Gravity} (GQG), whose Lagrangian is a sum of the Gauss-Bonnet and QG Lagrangians. Our analysis of causality constraint in GQG has two key components: to construct an exact shock wave solution and to then study the Shapiro time shift of a massless probe graviton crossing it. 

We will show that the presence of massive spin-2 ghost modes can restore causality in GQG for a non-trivial range of classical couplings. As the existence of propagating ghosts challenges the unitarity of theories like QG and GQG, the restoration of causality due to such modes thus helps us in distilling various consistency constraints used to classify viable effective theories of gravity.\footnote{This is reminiscent of what happens in $3D$ gravity, where causality is in line with unitarity and at odds with a well-posed black hole spectrum \cite{Edelstein:2016nml}.} For example, in $D=5$ dimensions, we show that there is a classical regime of parameter space in which causality is respected for the eikonal scattering of massless gravitons in the shock wave geometry.

Following CEMZ's calculation in EGB gravity~\cite{CEMZ}, one may be tempted to conclude that any theory (like GQG) whose three-point graviton coupling around flat spacetime differs from that of GR would be causality violating. However, our computation of the Shapiro time shift in GQG theory demonstrates that is not the case. In fact, even though EGB theory violates causality, a non-trivial deformation of the Lagrangian by adding QG terms can restore it. We also emphasize that our results are not in conflict with the CEMZ criterion as the shock wave solution of GQG is not analytically related to that of GR and the propagating modes in GQG include a massive spin-$2$ ghost.

We emphasize that GQG, owing to the presence of massive perturbative ghosts, may not be regarded as a consistent or physically viable theory of gravity. These ghost degrees of freedom generate perturbative instabilities, thereby undermining the theory’s status as a well-defined framework. Yet, as we shall demonstrate, it is precisely the presence of these ghost modes that enforces the CEMZ causality criterion. Thus, our analysis brings out an intriguing interplay: the presence of ghosts, while being a potential source of instability, also appears to be an unexpected possible remedy to causality violation. While the ultimate status of GQG is yet to be settled, our work represents an initial step toward addressing it.

\section{Causality constraint and CEMZ criterion}
Gao and Wald proposed a notion of causality in curved spacetime \cite{GaoWald}, which states that the asymptotic structure sets a lower limit on the time taken by any signal to travel from one detector at spatial infinity to another. In GR, with matter satisfying null energy condition, this criterion is always satisfied in four dimensions. As argued by CEMZ, this proof can be generalised beyond GR and for higher dimensions as well. CEMZ regarded the Gao-Wald notion of causality as a requirement for any theory of gravity and used it to constrain higher curvature terms. Such constraints stem from the fact that since the Gao-Wald criterion does not simply depend on the asymptotic causal structure alone, higher curvature terms do not unconditionally respect this notion of causality. 

CEMZ showed that in any theory of gravity, the Gao-Wald criterion transmutes into the sign of the Shapiro time shift suffered by a probe graviton scattering through a shock wave with negligible backreaction.\footnote{If the magnitude of the spatial linear momentum of the shock wave is $p$ and of the probe graviton $p^{\prime}$, then backreaction will be negligible for $p^{\prime}\, \ll \, p$.} For theories in which the probe graviton suffers a time advancement, it would be possible to send a signal between two detectors placed at infinity faster through the bulk than asymptotically, violating the Gao-Wald criterion of causality. CEMZ used this diagnostic to show that any higher curvature theory of gravity that leads to a distinct three-graviton interaction from GR for perturbations around flat space must necessarily be causality violating.

Note that the CEMZ analysis can be applied in $D = 4$ dimensions with two important caveats. The Shapiro time delay is logarithmic and thereby depends on the choice of an infrared cutoff. Most importantly for our current purpose, the quadratic Riemann contribution is irrelevant in four dimensions due to the Gauss-Bonnet theorem. Therefore, in order to perform the analysis of this paper in $D = 4$, we would need to deal with cubic gravities, certainly an interesting but more involved case. Instead, we will work in $D > 4$ and study the simpler quadratic case. For more details on the $D = 4$ scenario, we refer the reader to Ref.~\cite{Camanho:2015bbt,Hollowood:2016ryc}.

A negative Shapiro time shift could theoretically be used to construct closed timelike curves. Although this could cause a significant violation of macroscopic causality, recent studies have cast doubt on these constructions \cite{Papallo:2015rna, Kaplan:2024qtf}. Nevertheless, there are other potential pathological consequences associated with a negative Shapiro time shift. If we consider the process of a probe graviton scattering through a shock wave as a signal transmission problem, causality demands that the $S$-matrix element for graviton-graviton scattering amplitudes must satisfy the condition $|S(\omega)| \leq 1$ for Im$(\omega) > 0$, which would be violated if the probe graviton experiences a negative Shapiro time shift~\cite{CEMZ}.

Based on this criterion, CEMZ has explicitly demonstrated that EGB gravity is acausal and must be completed with an infinite tower of massive higher spin particles to restore causality. Then, is it a generic issue in any higher curvature theory of gravity beyond GR? It turns out that the answer to this important question is negative. In fact, in a previous work~\cite{EGLS}, we showed that the QG theory is not causality violating and cannot be ruled out by the CEMZ criterion despite the fact that QG is not equivalent to GR without matter via any field redefinition. The subsequent section delves into further generalizations of CEMZ's result by incorporating a new class of theories in higher dimensions ($D>4$).

Let us emphasize that our motivation behind considering higher dimensions is to understand whether the time advancement arising due to the Gauss-Bonnet term can be cured by adding other higher curvature terms with non trivial graviton 3-point coupling. Note that the Gauss-Bonnet term is trivial in $D=4$. This is why we primarily focused on the higher dimensions and analyze whether using the extra terms along with Gauss-Bonnet can make it possible to find a classical parameter space where the Shapiro time delay remains positive.

\section{Shapiro time shift for GQG}
The analysis of Ref.~\cite{EGLS} reveals that relating causality with deformation of three and higher point couplings is rather subtle in theories with massive ghosts. In light of this realisation,  we now consider the simplest $D \geq 5$ case, Generalised Quadratic Gravity, described by the Lagrangian
\begin{equation}
{\cal L}_{\text{GQG}} = \frac{1}{16 \pi G} \left[ R + \alpha \, R^2 + \beta \, R_{ab} R^{ab} + \gamma \, {\cal L}_{\text{GB}} \right] ~,
\end{equation}
where ${\cal L}_{\text{GB}} = R^2 - 4 R_{ab} R^{ab} + R_{abcd}R^{abcd}$ is the Gauss-Bonnet term. For perturbations around a flat background, it may lead to tachyonic instabilities unless $\beta \leq 0$, and $4\, (D-1)\, \alpha + D \, \beta \geq 0$~\cite{Tekin:2016vli}. 

Even in the absence of tachyons, GQG theory does however contain massive spin-2 ghosts that may violate linearized stability. Nevertheless, as mentioned in the Introduction, the ultimate fate of gravitational theories with ghosts is subtle and remains an active area of investigation. Moreover, there are indications that at least some of the ghost-induced pathologies could be an artifact of the perturbative analysis. This claim is supported by the fact that these theories admit well-defined conserved charges, with an energy expression that remains positive in a suitable region of the couplings \cite{DT2002, DT2003}. In fact, it was recently proved that some higher curvature theories have a well posed initial value formulation when their Lagrangian is augmented by suitable regularising terms \cite{Figueras:2024bba}. 

At this point, we must also emphasize that our primary motivation for investigating higher curvature theories like GQG is to explore the status of the CEMZ causality criterion beyond the framework of GR. Unlike EGB gravity, where a violation of the CEMZ criterion is well-known, it has been observed that QG respects this requirement~\cite{EGLS}. This observation naturally prompts us to study a more general class of higher curvature theories, with GQG serving as a simple yet compelling candidate. Also, as mentioned before, a particularly noteworthy feature of GQG is that its three-point graviton interaction vertex differs non-trivially from that in GR, offering a richer structure to probe. Consequently, GQG not only broadens our understanding of the CEMZ framework but also sheds light on how different curvature corrections influence causal propagation.

However, it is important to underscore a significant caveat associated with higher curvature theories: the ubiquitous presence of perturbative ghosts, which generally threatens the stability and viability of the theory. Unless it can be convincingly demonstrated that such ghosts are rendered benign beyond the perturbative regime--possibly via non-perturbative effects or a reformulation of the theory--the underlying theory must be deemed unphysical in the conventional sense. That said, our current objective is not to advocate for the physical viability of GQG or similar higher curvature models. Rather, we aim to systematically investigate the interplay between various physical criteria, including perturbative stability, the emergence of ghost modes, and compliance with the CEMZ causality condition. Through such analysis, we hope to probe the extent to which these conditions are interlinked or mutually constraining within the broader landscape of modified gravity theories.

\subsection{Constructing the shock wave solution}
A rather non-trivial aspect of GQG is that it supports an exact shock wave solution that is not analytically related to the GR shock wave. This can be seen as follows. The general form of a shock wave in terms of coordinates $\{u, v, x_i\}$ is given by \cite{CEMZ}, 
\begin{align}
\dd s^2 = - \dd u\, \dd v + h_{0}(u,x_i)\, \dd u^2 + \delta_{ij}\, \dd x^{i}\, \dd x^{j}.
\end{align}
Here, $(u,v)$ are the pair of null coordinates and $x_i$'s are the remaining $(D-2)$-transverse coordinates. To find a shock wave solution, we consider the following ansatz~\cite{CEMZ}, 
\begin{align} \label{h0}
h_{0}(u,x_i)\, =\, {\cal F}(r)\, \delta(u)\, ,
\end{align}
with $r=\sqrt{x_i\, x^i}$ being the transversal distance from the shock confined in the $u=0$ plane.

Rather interestingly, the shock wave solution in GQG can be identified with shock wave solution in QG as ${\cal L}_{\text{GB}}$ does not contribute to the field equation~\cite{CEMZ}. Hence, the shock-profile of GQG is given by
\begin{align} \label{fr}
{\cal F}(r) = \frac{4\, G\, \vert P_{u}\vert\, \Gamma(n)}{\pi^{n}\, r^{2n}}\, \left[1 - E_n(x)\right],
\end{align}
where $P_u$ is the momentum of the probe particle moving along the $v$-direction, the function $E_n(x)$ is related to the modified Bessel function of 2nd kind $K_{-n}(x)$ as $E_n(x) = \left[2^{n-1}\Gamma(n) \right]^{-1}\, x^n\, K_{-n}(x)$ with $n = (D-4)/2$ and $x=r/\sqrt{-\beta}$. We note that $\beta \leq 0$ is a necessary condition for the absence of tachyons. Just as in QG, ${\cal F}$ is independent of $\alpha$. However, we must choose a non-zero and positive value of $\alpha$ to ensure that the perturbations around flat space do not contain a tachyon. Now, two comments are in order: 

\noindent
(i) It can then be immediately verified that for $\beta \to 0^-$, this shock wave reduces to the $D$-dimensional GR solution that is also the shock wave solution of EGB gravity for any $D \geq 5$. \\

\noindent
(ii) However, this shock wave solution cannot be analytically continued to that of GR~\cite{EGLS}. Thus, the shock wave geometry of GQG is not in the same ``universality'' class as that of GR and no continuous flow in the space of couplings connects the two solutions. 

\subsection{Shapiro time shift computation}
In order to subject classical GQG theories to the CEMZ diagnostic, we now consider the scattering of a probe graviton in the above shock wave background. The impact parameter in the transversal space will be denoted by $\vec{b}$ with its magnitude being denoted by $b$. That is, we consider a massless spin-2 plane wave traveling in $u$-direction which scatters through the shock wave at a finite spatial separation $b$ and propagates as a free graviton after crossing the shock wave at $u = 0$. The relevant graviton perturbation equation in the transverse traceless (TT) gauge \footnote{In the shock wave background, the relevant degrees of freedom of the spin-2 perturbation $h_{\mu\nu}$ are in the TT sector alone, provided we assume an auxiliary condition on the polarisation tensor: $\epsilon_{ij}\, b^i\, b^j=0$. For more details, the reader can consult Sec. IV-(A) of Ref.~\cite{Benakli:2015qlh}.} is as follows,
\begin{equation} \label{pert}
    \left( \Box + \beta\,  \Box^2\right) h_{ij} - 16\, \gamma \left( \partial_{(i|}\, \partial_k\, h_0\right) \partial_v^2\, h_{k |j)} = 0\, ,
\end{equation}
where $\Box = -4 \left(\partial_u \partial_v + h_0\, \partial_v^2 \right)$ is the d'Alembert operator in the shock wave spacetime. This equation leads to a solution for $\partial_v h_{ij} = \bar{h}_{ij}$ of the form: 
\begin{align}
\bar{h}_{ij} \sim \, \epsilon_{ij}\, \int dp_v\, \exp \left[i\, p_{v} \left\{ v - \int_{u} \left(h_{0} + \gamma\, h_{1}\right) du \right\} \right]\, , \nonumber
\end{align}
where $h_0$ is given by Eq.~\eqref{h0} and $\epsilon_{ij}$ is the polarisation tensor of the probe graviton. Note that our analysis is valid as long as the probe graviton can be treated as a test particle in the fixed shock wave background $g_{\alpha \beta}$. At the level of metric perturbation $h_{ij}$, this is guaranteed if we are in the domain of linear perturbation theory satisfying $\text{max}[|h_{ij}|] \ll |g_{\alpha \beta}| \sim 1$.

Owing to the type-$N$ (in the Petrov classification) nature of the shock wave background, all non-vanishing curvature components are proportional to $k_\mu k_\nu$, where $k^\mu$ is the principal null direction. This severely restricts the possible curvature-coupling terms in the linearized equations and strongly suggests that the massless and massive spin-2 sectors remain decoupled during scattering. Indeed, in the eikonal regime characterized by large impact parameters and small momentum transfer, an incoming massless graviton can only scatter into outgoing massless gravitons and cannot produce massive spin-2 ghost states. The reason is kinematical: producing an on-shell ghost of mass $M_g$ requires a longitudinal momentum transfer of order $M_g^2/E$, where $E$ is the center-of-mass energy of the incoming graviton, so all inelastic transitions are suppressed by powers of $1/E$ relative to elastic scattering in the eikonal limit $E \to \infty$ at fixed $b$. At leading eikonal order, the scattering is therefore elastic and diagonal in particle species, and the production of massive spin-2 ghosts is kinematically forbidden. Hence, we impose an asymptotic condition on the perturbation that it is a free massless spin-2 graviton before and after crossing the shock wave. As a result, $h_{1}$ must be of the form $h_{1} := {\cal G}(x_i)\, \delta(u)$. We can thus arrive\footnote{Here, we have used $\int_{-\epsilon}^{+\epsilon} du\, \delta^2(u) = 0$. A naive way to derive this identity would be to use point-splitting and interpret $\delta^{2}[f] :=\, \lim_{\epsilon\, \rightarrow\, 0}\, \int du\, \delta(u + \epsilon) ( \delta (u - \epsilon)\, f(u) )$ for any integrable function $f$. A more rigorous way can be found in Refs. \cite{delta_ref1, delta_ref2}.} at the following expression of the Shapiro time shift as the probe graviton crosses the shock with an impact parameter $r=b$,
\begin{equation}
\begin{split}
(\Delta v)_{\text{GQG}} & = \left(\Delta v\right)_{\text{EGB}} - \left[ (1 + 4 \eta \hat{\gamma}) \, E_n(x_b) \right. \\
& \left. + 8 n \hat{\gamma} x_b^{-2} \left\{ 2(n+1) \eta - 1 \right\}\, E_{n+1}(x_b) \right] \left(\Delta v\right)_{\text{GR}} ~,
\end{split}
\label{dv}
\end{equation}
where we have used the notations $\hat{\gamma}=\gamma/|\beta_e|$, and $x_b = x(r=b)$. Moreover, $\left(\Delta v\right)_{\text{GR}}$ is the Shapiro time shift for GR and $\left(\Delta v\right)_{\text{EGB}}$ is that for EGB gravity~\cite{CEMZ},
\begin{equation}
\left(\Delta v\right)_{\text{EGB}} = \left[ 1 + 8 n \hat{\gamma} x_b^{-2} \left\{ 2(n+1) \eta - 1 \right\} \right] \left(\Delta v\right)_{\text{GR}} ~.
\end{equation}
Here, $\eta := (\epsilon \cdot \hat{b})^{2}/(\epsilon \cdot \epsilon)$ is the normalized projected polarisation tensor (with $\hat{b}$ is the unit vector along $\vec{b}$) and we have $0 \leq \eta \leq 1 / 2$. This allowed range of $\eta$ can be derived as follows. For rotational invariance in the $(D-2)$-dimensional flat transversal plane, it is always possible to set $\hat{b} = (1,0,0,\cdots)$. Then, due to the auxiliary condition $\epsilon_{ij} b^i b^j =0$, we get $\epsilon_{11}=0$. Hence, $\eta = (\epsilon^{1i}{\epsilon_{1i}})/(2 \epsilon^{1j}{\epsilon_{1j}} +\text{non-negative\, terms}) \in [0,1/2]$.

One can also straightforwardly check that Eq.~\eqref{dv} leads to correct GR and EGB limits. For this purpose, it suffices to note that the function $E_n(x_b)$ goes to zero as $x_b \to \infty $, \textit{i.e.}, $\beta \to 0^-$. Moreover, when $\hat{\gamma} \to 0$, the Shapiro time shift coincides with the QG result in Ref.~\cite{EGLS}.

\section{Shapiro time Shift in $D=5$ dimensions}
We would like to analyse whether, like in the case of EGB gravity, $(\Delta v)_{\text{GQG}}$ can flip its sign for a choice of the probe polarisation in $5$-dimensions. Since we are working within the eikonal regime, the impact parameter satisfies $b \gg 1/p_{v}$. We consider the domain of dimensionful couplings in the ${\cal O}(1)$ neighbourhood of $\gamma \sim \beta$ and choose $b^{2}\, \sim\, |\beta|$. This implies that $\hat{\gamma} = {\cal O}(1)$. Then, to ascertain the sign of the Shapiro time shift, we consider  $b^{2} = \gamma = -\beta/k$ with $k \gtrsim {\cal O}(1)$. We remind the reader that $k$ is positive due to the tachyon-free condition.

Note that, as in the original CEMZ analysis of EGB gravity \cite{CEMZ}, the regime of interest lies at impact parameters of the order of the scale set by the higher-curvature couplings, $b \sim M_g^{-1}$. Our statements therefore concern GQG regarded as an exact classical theory with finite couplings, rather than as a truncated effective expansion in Planck units, and it is the exactness of the shock wave solution \eqref{fr} that renders this regime accessible.

We now search the $(\eta, k)$ space and find the region for which the time shift is positive for the entire range of the polarisation. 
\begin{figure}[ht!]
\centering
\includegraphics[width=0.7\columnwidth]{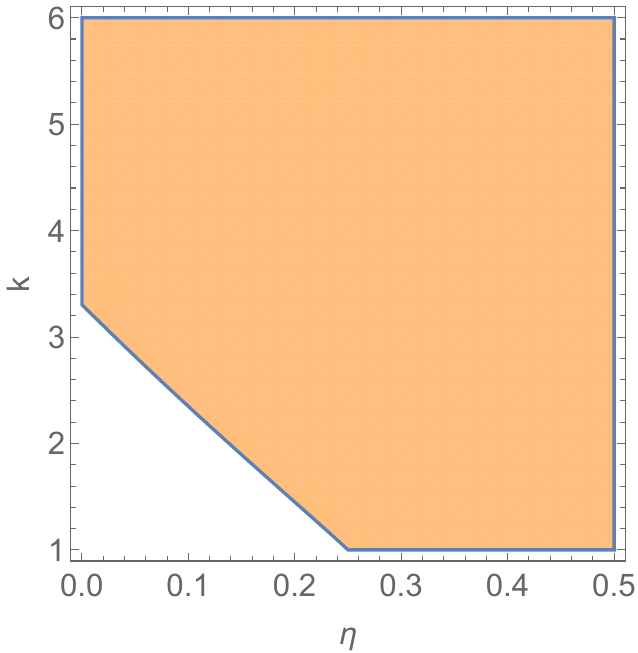}
\caption{Variation of the Shapiro shift in $D=5$ w.r.t $(\eta,\, k)$ for $\gamma = b^2$. The orange shaded region is where $(\Delta v)_{\text{GQG}} > 0$.} 
\label{fig:D5}
\end{figure}
The result is presented in Fig.~[\ref{fig:D5}], which clearly shows the existence of a region within the allowed values of $\eta$, where $k \gtrsim {\cal O}(1)$ and $(\Delta v)_{\text{GQG}} > 0$. More generally, setting $b^{2} = \gamma/m$, one can explicitly check that the Shapiro time shift is always positive irrespective of the graviton polarisation if $k \gtrsim 4\, m^2$ (for $m \geq 0$) and $k \gtrsim {\cal O}(1)$ (for $m \leq 0$). Hence, for such choices of parameters, there is no violation of causality in the sense of CEMZ.

To intuitively understand these results, we refer to the expression given in Eq.~\eqref{dv}. By choosing a polarisation $\eta$ that makes $\left(\Delta v\right)_{\text{EGB}} < 0$, we find that there exists a critical value of $\beta$ beyond which the second term in Eq.~(\ref{dv}) becomes dominant, resulting in an overall positive time shift $(\Delta v)_{\text{GQG}}$. Achieving this within the classical regime indicates the restoration of causality in GQG. Therefore, given the space $S_{\textrm{EGB}}$ of classical couplings in EGB theory that violates causality, there exists a subspace $S_{\textrm{GQG}}$ of GQG couplings ($S_{\textrm{GQG}} \supset S_{\textrm{EGB}}$) with non-zero values $\beta$ where there is no time advancement for any allowed choice of the graviton polarisation and impact parameter $\vec{b}$. Notice that for large $k$ the impact parameter is smaller than the Compton wavelength of the ghost massive graviton; the probe is close enough to feel the virtual ghost via its Yukawa tail, and this seems to be at the root of causality restoration. Yet, it can not produce the ghost because such a process is blocked by longitudinal kinematics in the eikonal regime, as discussed above. Therefore, even though EGB theory violates causality, a deformation of the EGB Lagrangian by QG terms can restore causality in the regime of the classically allowed parameter space  due to the presence of the massive ghost mode.

\section{Shapiro time shift in $D \geq 6$ dimensions}
Given that the behavior of the shock wave profile in Eq.~\eqref{fr} is highly sensitive to $D$, we anticipate that the Shapiro time shift will exhibit certain interesting features as $D$ varies. Specifically, for $D=6$ and a particular choice of $(m,\eta)$, numerical inspections reveal that for the Shapiro time shift to remain positive, $k$ must satisfy $k \gtrsim 0.29\, \exp[8m(1-2 \eta)]$. Notably, for positive values of $m$, exponentially large values of $k$ are required to ensure that $(\Delta v)_{\text{GQG}} > 0$ for all $\eta \in [0,1/2]$. Conversely, for negative $m$ ({\it i.e.}, $\gamma < 0$), any ${\cal O}(1)$ or greater values of $k$ would suffice. Thus, in six dimensions, the parameters of the GQG theory can be chosen such that it remains a classically viable theory that respects causality.

However, we shall now see that the same is not possible as we increase the values of $D$. In fact, for $D \geq 7$, a Taylor expansion in $1/k$ for $k \gg 1$ shows that to maintain causality the values of $m$ must lie within the range $(0,1/\{4(D-6)\})$ or $(-1/\{2(D-6)^2\},0)$. For any values of $m$ outside this range, it can be checked numerically that $(\Delta v)_{\text{GQG}}$ cannot be kept positive with any allowed values of $k$ in $D > 6$.
\begin{figure}[ht!]
\includegraphics[width=\linewidth]{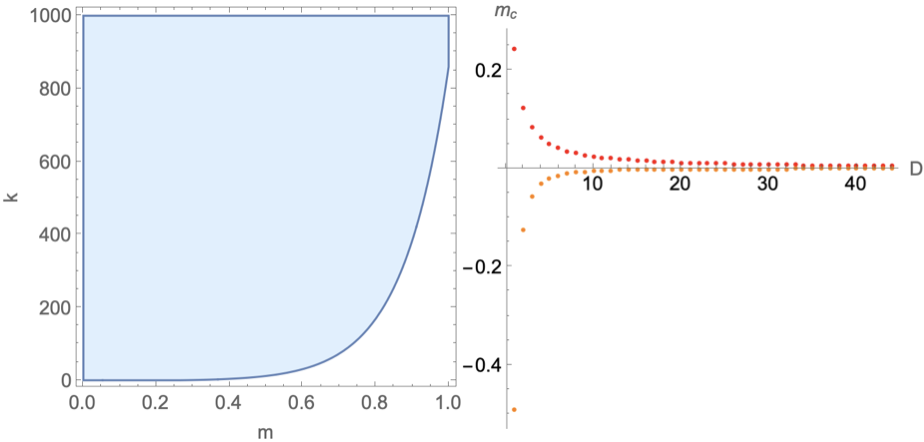}
\caption{Variation of the Shapiro shift in $D = 6$ for $m \in (0,1]$ and $k \in [1,1000]$ with $\eta = 0$ (left panel). The shaded regions are where $(\Delta v)_{\text{GQG}} > 0$. The corresponding plot for negative values of $m$ and $\eta = 1/2$ does not put extra constraints on the allowed values of $k$. The right panel plots the critical values $m=m_c$ (outside which time shift becomes negative) as a function of $D \in [7,50]$ for $\eta=0$ (red dots) and $\eta=1/2$ (orange dots) with $k = 1000$.} 
\label{fig:D6}
\end{figure}

To illustrate this surprising feature, we have shown in the left panel of Fig.~\ref{fig:D6}, the variation of the Shapiro time shift with respect to $(m,k)$ at $D = 6$ for the cases with $\eta = 0$. The right plot demonstrates that as $D$ increases, the critical values of $m$ approach zero, resulting in a progressively higher required value of $\gamma$. Consequently, restoring causality within the classically allowed parameter space becomes increasingly challenging. In fact, in the limit $D \to \infty$ with a fixed $k$, it can be verified that $(\Delta v)_{\text{GQG}} \to -(2 m/3)\, (\Delta v)_{\text{GR}}$ for $\eta = 0$, and $(\Delta v)_{\text{GQG}} \to (2 m n/3)\, (\Delta v)_{\text{GR}}$ for $\eta = 1/2$. As a result, the critical values of $m$ tends to zero as $D \to \infty$, ruling out the existence of ``causality preserving'' regions in the parameter space of classical couplings. Although the underlying physical reasoning behind such effects remains elusive to us at present, we are hopeful that future research in this direction may shed light on aspects such as, for instance, their meaning in the context of the large $D$ limit of GR \cite{Emparan:2013moa}.

\section{Discussion and Conclusions}
It is crucial to explore theories beyond GR that satisfy all consistency criteria. For instance, if we demand a gravity theory be free from perturbative ghosts and tachyons, Lovelock gravity emerges as the sole option. However, as demonstrated by CEMZ~\cite{CEMZ}, Lovelock theories with a non-vanishing EGB coupling exhibit causality issues, resulting in Shapiro advancements and the violation of the CEMZ causality criterion. Consequently, these theories can be ruled out as standalone classical theories of gravity.

In this work, we demonstrated that the situation differs significantly for higher curvature theories like GQG, which introduces a massive graviton mode as a perturbative ghost and features a different graviton three-point coupling compared to GR. Unlike EGB gravity, the Shapiro time shift in GQG remains strictly positive within a classically allowed parameter space, avoiding causality issues per the CEMZ criterion. Although it was pointed out in Ref.~\cite{CEMZ} that a massive graviton might restore causality, the possibility was dismissed there on the basis of an argument involving angular momentum conservation, which seems evaded in GQG by its ghosty nature.

What do these results imply for the consistency of higher-curvature theories beyond GR? First, we find that the inclusion of higher-curvature terms does not necessarily lead to a violation of the CEMZ causality requirement. This is exemplified by GQG, which admits modified graviton three-point couplings while still satisfying the CEMZ criterion. Nevertheless, the issue of perturbative instability due to ghost modes persists in GQG, and in this sense the theory still cannot be regarded as a healthy theory of gravity. However, recent studies of certain systems containing ghost modes suggest that instabilities may be evaded under special structural conditions \cite{Deffayet:2023wdg} (see also \cite{Held:2023aap, Figueras:2024bba, Held:2025ckb}). Whether any such mechanism extends to quadratic gravity ---let alone to GQG, which would require generalizing the analysis of \cite{Deffayet:2023wdg} to include $Riemann^2$ terms--- remains open: our analysis does not, by itself, establish that GQG is stable or well-behaved. In light of our result demonstrating the validity of the CEMZ condition in GQG, pursuing such a generalization would be of considerable interest.

Moreover, an intriguing connection between the CEMZ criterion and $S$-matrix analyticity bounds has been proposed in Ref.~\cite{Caron-Huot:2022ugt}, arguing that causality constraints can be recast in terms of analyticity/positivity properties of scattering amplitudes. In this context, our results raise an interesting and nontrivial question: whether the violation of unitarity (due to ghost modes) and violations of micro-causality (due to higher-curvature terms like Gauss–Bonnet) might interplay in such a way that the resulting effective field theory still satisfies $S$-matrix analyticity bounds. A detailed investigation of this issue lies beyond the scope of the present work. In particular, it would require a careful analysis of scattering amplitudes in ghost-containing theories such as GQG and their analyticity properties. We leave this as an interesting direction for future study.

Irrespective of the validity of GQG as a viable theory of gravity, our study isolates unitarity issues from the causality constraints as it finds an allowed region of parameter space where the theory is causal despite the threat of the EGB three-point couplings. There is nothing in our analysis that seems special of the quadratic case; it should extend smoothly to the cubic scenario, albeit the greater abundance of terms might make the analysis significantly more involved. This is certainly a problem we hope to solve in the near future.

In conclusion, our results pave the way for a deeper understanding of higher curvature gravity, indicating that GQG-like theories can meet the essential consistency requirement of CEMZ causality and potentially other criteria as well. This may open new avenues for exploring the intricate balance between causality and ghost stability within the broader landscape of beyond-GR theories.

\section*{Acknowledgements}
J.D.E. is supported by AEI-Spain PID2023-152148NB-I00 and by Maria de Maeztu excellence unit grant CEX2023-001318-M, by Xunta de Galicia (CIGUS Network of Research Centres, projects ED431C-2021/14 and ED431F-2023/19), and by the European Union FEDER. R.G. is supported by the Fulbright Nehru Postdoctoral Research Fellowship (Award No.3174/ FNPDR/2025) from the United States-India Educational Foundation. R.G. acknowledges the warm hospitality at IIT Gandhinagar, where a part of this work was completed. The research of S.S. is funded by the Department of Science and Technology, Government of India, through the SERB CRG Grant (No. CRG/2023/000545). S.S. also extends his gratitude to the Instituto Galego de Física de Altas Enerxías (IGFAE) in Spain for their excellent hospitality during his sabbatical leave, where a part of this work was completed.


\begin{thebibliography}{100}
 
\bibitem{Will:2014kxa}
C.~M.~Will,
Living Rev. Rel. \textbf{17}, 4 (2014).

\bibitem{Berti:2015itd}
E.~Berti, E.~Barausse, V.~Cardoso, L.~Gualtieri, P.~Pani, \textit{et al.}
Class. Quant. Grav. \textbf{32}, 243001 (2015).

\bibitem{Berti:2019tcy}
E.~Berti,
Gen. Rel. Grav. \textbf{51}, 140 (2019).

\bibitem{Stelle:1977ry}
K.~S.~Stelle,
Gen. Rel. Grav. \textbf{9}, 353 (1978).

\bibitem{Audretsch:1993kp}
J.~Audretsch, A.~Economou and C.~O.~Lousto,
Phys. Rev. D \textbf{47}, 3303 (1993).

\bibitem{Salvio:2018crh}
A.~Salvio,
Front. in Phys. \textbf{6}, 77 (2018).

\bibitem{Padmanabhan:2013xyr}
T.~Padmanabhan and D.~Kothawala,
Phys. Rept. \textbf{531}, 115 (2013).

\bibitem{Camanho:2013pda}
X.~O.~Camanho, J.~D.~Edelstein and J.~M.~S\'anchez de Santos,
Gen. Rel. Grav. \textbf{46}, 1637 (2014).

\bibitem{Bueno:2016ypa}
P.~Bueno, P.~A.~Cano, V.~S.~Min and M.~R.~Visser,
Phys. Rev. D \textbf{95}, 044010 (2017).

\bibitem{Bueno:2019ltp}
P.~Bueno, P.~A.~Cano, J.~Moreno and \'A.~Murcia,
JHEP \textbf{11}, 062 (2019).

\bibitem{Deffayet:2023wdg}
C.~Deffayet, A.~Held, S.~Mukohyama and A.~Vikman,
JCAP \textbf{11}, 031 (2023).

\bibitem{Held:2023aap}
A.~Held and H.~Lim,
Phys. Rev. D \textbf{108}, 104025 (2023).

\bibitem{Figueras:2024bba}
P.~Figueras, A.~Held and \'A.~D.~Kov\'acs,
arXiv:2407.08775 [gr-qc].

\bibitem{Held:2025ckb}
A.~Held and H.~Lim,
arXiv:2503.13428 [gr-qc].

\bibitem{Cline:2003gs}
J.~M.~Cline, S.~Jeon and G.~D.~Moore,
Phys. Rev. D \textbf{70}, 043543 (2004).

\bibitem{Antoniadis:1986tu}
I.~Antoniadis and E.~T.~Tomboulis,
Phys. Rev. D \textbf{33}, 2756 (1986).

\bibitem{Smilga:2005gb}
A.~V.~Smilga,
Phys. Lett. B \textbf{632}, 433 (2006).

\bibitem{CEMZ}
X.~O.~Camanho, J.~D.~Edelstein, J.~Maldacena and A.~Zhiboedov,
JHEP \textbf{02}, 020 (2016).

\bibitem{EGLS}
J.~D.~Edelstein, R.~Ghosh, A.~Laddha and S.~Sarkar,
JHEP \textbf{09}, 150 (2021).

\bibitem{Edelstein:2016nml}
J.~D.~Edelstein, G.~Giribet, C.~Gomez, E.~Kilicarslan, M.~Leoni and B.~Tekin,
Phys. Rev. D \textbf{95}, 104016 (2017).

\bibitem{GaoWald}
S.~Gao and R.~M.~Wald,
Class. Quant. Grav. \textbf{17}, 4999 (2000).

\bibitem{Camanho:2015bbt}
X.~O.~Camanho, J.~D.~Edelstein and A.~Zhiboedov,
Int. J. Mod. Phys. D \textbf{24}, 1544031 (2015).

\bibitem{Hollowood:2016ryc}
T.~J.~Hollowood and G.~M.~Shore,
J. Phys. A \textbf{49}, 215401 (2016).

\bibitem{Papallo:2015rna}
G.~Papallo and H.~S.~Reall,
JHEP \textbf{11}, 109 (2015).

\bibitem{Kaplan:2024qtf}
D.~E.~Kaplan, S.~Rajendran and F.~Serra,
arXiv:2406.06681 [hep-th].

\bibitem{Tekin:2016vli}
B.~Tekin,
Phys. Rev. D \textbf{93}, 101502 (2016).

\bibitem{DT2002}
S.~Deser and B.~Tekin,
Phys. Rev. Lett. \textbf{89}, 101101 (2002)

\bibitem{DT2003}
S.~Deser and B.~Tekin,
Phys. Rev. D \textbf{67}, 084009 (2003)

\bibitem{Benakli:2015qlh}
K.~Benakli, S.~Chapman, L.~Darm\'e and Y.~Oz,
Phys. Rev. D \textbf{94}, 084026 (2016).

\bibitem{delta_ref1}
E.~L.~Koh, and L.~C.~Kuan,
Math. Nachr. \textbf{157}, 243 (1992).

\bibitem{delta_ref2}
C.~K.~Li and C.~P.~Li,
Appl. Math. Comput. \textbf{246}, 502 (2014).

\bibitem{Emparan:2013moa}
R.~Emparan, R.~Suzuki and K.~Tanabe,
JHEP \textbf{06}, 009 (2013)


\bibitem{Caron-Huot:2022ugt}
S.~Caron-Huot, Y.~Z.~Li, J.~Parra-Martinez and D.~Simmons-Duffin,
JHEP \textbf{05}, 122 (2023).


\end{thebibliography}
\end{document}